\documentclass[preprintnumbers, aps, floatfix, twocolumn, preprintnumbers, letterpaper, superscriptaddress,nofootinbib,10pt]{revtex4-2}
\usepackage{multirow,dcolumn}

\usepackage{graphicx}
\usepackage{caption}
\usepackage{epsfig}
\usepackage{bm}
\usepackage{amsfonts}
\usepackage[T1]{fontenc}
\usepackage{subfigure}
\usepackage{verbatim}
\usepackage{amssymb}
\usepackage{color}
\usepackage{float}
\usepackage{amsmath}
\usepackage{urwchancal}
\usepackage{dcolumn}
\usepackage[normalem]{ulem}
\usepackage{cancel}
\usepackage[colorlinks]{hyperref}
\usepackage{xcolor}
\hypersetup{
     breaklinks=true,
    pdfstartview={FitH},    
    colorlinks=true,       
    linkcolor=blue,          
    citecolor=red,        
    filecolor=magenta,      
    urlcolor=blue,           
    anchorcolor=green,      
    linktocpage=true
}
\usepackage{microtype}

\begin{document}

\title{Lyapunov Exponents and Phase Transitions in Four-Dimensional AdS Black Holes with a Nonlinear Electrodynamics Source}

\author{Ram\'on B\'ecar}
\email{rbecar@uct.cl} \affiliation{\small{Departamento de Ciencias Matem\'aticas y F\'{i}sicas, Universidad Cat\'olica de Temuco, Montt 56, Casilla 15-D, Temuco, Chile.}}
\author{P. A. Gonz\'{a}lez}
\email{pablo.gonzalez@udp.cl} \affiliation{\small{Facultad de
Ingenier\'{i}a y Ciencias, Universidad Diego Portales, Avenida Ej\'{e}rcito
Libertador 441, Casilla 298-V, Santiago, Chile.}}
\author{Felipe Moncada}
\email{fmoncada@uct.cl} \affiliation{\small{Departamento de Ciencias Matem\'aticas y F\'{i}sicas, Universidad Cat\'olica de Temuco, Montt 56, Casilla 15-D, Temuco, Chile.}}
\author{Yerko V\'{a}squez}
\email{yvasquez@userena.cl}\affiliation{\small{Departamento de F\'{\i}sica, Facultad de Ciencias, Universidad de La Serena,\\ 
Avenida Cisternas 1200, La Serena, Chile.}}
\date{\today}

\begin{abstract}

We investigate the relationship between dynamical instability and thermodynamic phase transitions in four-dimensional Anti--de Sitter black holes in Einstein gravity coupled to a nonlinear power-law electromagnetic field with exponent $p = 3/4$. In the canonical ensemble, we identify a critical electric charge $Q_c$ separating a regime exhibiting a first-order small/large black-hole (SBH/LBH) phase transition from a regime with a single thermodynamically stable phase. 
For both massless and massive probes, the thermal profile of the Lyapunov exponent $\lambda(T)$ becomes multivalued in the SBH/LBH coexistence region and exhibits a finite discontinuity at the transition temperature. This jump vanishes continuously as $Q \to Q_c$, signaling the termination of the first-order transition at a second-order critical point. Near criticality, the Lyapunov discontinuity obeys a universal mean-field scaling law with critical exponent $1/2$. For massless probes, we further analyze the critical impact parameter $b_c$, which displays the same multivalued structure and critical behavior as the Lyapunov exponent.
We also demonstrate that the spinodal temperatures, defined by the extrema of the $T(r_h)$ curve where the heat capacity at fixed charge diverges, coincide with singular features in the Lyapunov exponent.
Our results identify the Lyapunov exponent as a unified dynamical probe capable of capturing both first-order phase coexistence and second-order critical behavior in black-hole thermodynamics.

\end{abstract}
\maketitle

\flushbottom



\tableofcontents

\newpage

\section{Introduction}

Phase transitions in gravitational systems have attracted considerable attention since the discovery of the Hawking--Page transition in the four-dimensional Anti--de Sitter (AdS) spacetime, which describes a transition between thermal AdS at low temperatures and an AdS black hole at high temperatures~\cite{HP}. Unlike asymptotically flat black holes, which are thermodynamically unstable because of their negative specific heat, AdS black holes exhibit a richer thermodynamic behavior: large AdS black holes have positive specific heat and are thermally stable, while small ones have negative specific heat and approach the Schwarzschild limit. The phase structure also depends on the horizon topology, since AdS black holes may have $(d-2)$--dimensional compact Einstein horizons of positive, zero, or negative curvature~\cite{Birmingham:1998nr}. In particular, Schwarzschild--AdS black holes with planar or hyperbolic horizons are thermally stable and do not exhibit a Hawking--Page transition, whereas spherical horizons admit a transition between thermal AdS and large stable black holes. Treating the cosmological constant as a thermodynamic pressure and its conjugate as the thermodynamic volume has led to an extended phase-space formulation of black-hole thermodynamics, revealing Van der Waals--like phase transitions and critical behavior~\cite{Kubiznak:2016qmn,Belhaj:2013cva,Zeng:2016aly}. This framework also admits a natural interpretation within the AdS/CFT correspondence, where the Hawking--Page transition corresponds to a confinement/deconfinement transition in the dual field theory~\cite{Witten:1998zw}. Dynamical imprints of these thermodynamic transitions have been identified through perturbative analyses, including signatures in quasinormal-mode spectra~\cite{Shen:2007xk,Liu:2014gvf}. Black-hole phase transitions have also been extensively explored in the context of holographic superconductors, where the condensation of scalar hair induces superconducting-like phase transitions~\cite{Gubser:2008px,Hartnoll:2008vx,Jing:2011vz,Zhao:2012cn,Roychowdhury:2012hp,Gangopadhyay:2013qza,Dey:2014xxa,Dey:2014voa,Lai:2015rva,Ghorai:2015wft,Liu:2015lit,Sheykhi:2016kqh}. Further classification of these transitions has been achieved using Ehrenfest’s scheme and free-energy constructions~\cite{Banerjee:2010da,Banerjee:2010bx,Banerjee:2011au,Banerjee:2011raa,Banerjee:2010ve}, while complementary geometric approaches have been proposed to probe their underlying microscopic structure~\cite{W,R,Ruppeiner:1995zz}.

A pioneering connection between Lyapunov exponents and black-hole phase transitions was proposed by Guo \textit{et al.}~\cite{Guo:2022kio}, who demonstrated that the Lyapunov exponent associated with the motion of particles and ring strings in Reissner--Nordstr\"om--AdS black holes becomes multivalued across the transition, exhibiting a discontinuous change that behaves as an order parameter with a critical exponent of $1/2$. This result established the Lyapunov exponent as a useful dynamical probe of black-hole thermodynamics. In a subsequent study~\cite{Yang:2023hci}, the analysis was extended to Born--Infeld AdS black holes, where the Lyapunov exponents were computed for both timelike and photon geodesics. It was shown that the phase transition can also be characterized by multivalued Lyapunov exponents, and that the discontinuous change near the critical point behaves as an order parameter with a universal critical exponent of $1/2$. More recently, the thermodynamic behavior of charged Gauss--Bonnet AdS black holes was investigated through the Lyapunov exponent~\cite{Lyu:2023sih}. It was found that the relationship between the Lyapunov exponent and the Hawking temperature accurately captures the features of the Small/Large phase transition and even the presence of a triple point. Moreover, the difference of Lyapunov exponents at the transition point behaves as an order parameter, reinforcing its role as a dynamical probe of black-hole thermodynamics and stability, see \cite{Kumara:2024obd,Hale:2024lzh,Du:2024uhd,Shukla:2024tkw,Gogoi:2024akv,Promsiri:2024hrl,Chen:2025xqc,R:2025gok,Awal:2025irl,Yang:2025fvm,Kumar:2025kzt,Guo:2025pit,Bezboruah:2025udi,Zhang:2025cdx,Ali:2025ooh}, for other spacetimes. On the other hand,
it was shown that the changes of the photon sphere radius and the minimum impact parameter can serve as order parameters for the small-large black hole phase transition with an universal exponent of $1/2$ near the critical point for any dimension $d$ of spacetime \cite{Wei:2017mwc},  see also \cite{ Han:2018ooi,NaveenaKumara:2019nnt, Kumar:2024sdg}.

In the present work, we extend these analyses to four-dimensional Einstein--Power--Maxwell AdS black holes, focusing on the physically well-motivated case of a sublinear electromagnetic field with power-law exponent $p=3/4$,  motivated by the existence of exact hairy black-hole solutions~\cite{O.:2016wcf} and its physical relevance within nonlinear electrodynamics. This sublinear regime ($p<1$) softens the electric field near the horizon and induces significant deformations of the Hawking temperature and Helmholtz free energy compared to the Maxwell case ($p=1$), resulting in a richer phase structure and a clearer separation between small and large black-hole branches. Nonlinear electrodynamics was originally introduced to remove the divergence of the self-energy of point charges~\cite{BI} and naturally arises in the low-energy limit of heterotic string theory~\cite{Kats:2006xp,Anninos:2008sj,Cai:2008ph}. Such theories play an important role in the construction of regular black-hole solutions~\cite{AyonBeato:1998ub,AyonBeato:1999rg,Cataldo:2000ns,Bronnikov:2000vy,Burinskii:2002pz,Matyjasek:2004gh} and have motivated extensive studies of black-hole and brane solutions sourced by nonlinear electromagnetic fields, particularly in asymptotically AdS spacetimes~\cite{Hendi:2010zz,Hendi:2010bk,Hendi:2010kv,Roychowdhury:2012vj,Cai:2004eh,Dey:2004yt,Aiello:2004rz,Hendi:2009sw,Hendi:2013dwa,Hendi:2015hoa,Hendi:2016dmh}. The thermodynamics of Einstein--Born--Infeld and power-law electrodynamics black holes has revealed rich phase structures, including locally stable small black holes and nontrivial critical behavior~\cite{Miskovic:2008ck,Gonzalez:2009nn,Hendi:2010zza,Hendi:2012um}, with further extensions to Gauss--Bonnet, Einstein--dilaton, and higher-dimensional frameworks~\cite{Sheykhi:2009pf,Hendi:2015xya,Dehghani:2006zi,Zangeneh:2015wia}. Nonlinear electrodynamics also significantly affects holographic superconductors and the associated AdS/CFT phase structure~\cite{Jing:2011vz,Zhao:2012cn,Roychowdhury:2012hp,Gangopadhyay:2013qza,Dey:2014xxa,Dey:2014voa,Lai:2015rva,Ghorai:2015wft,Liu:2015lit,Sheykhi:2016kqh}. In particular, generalized Maxwell theories such as the Power--Maxwell Invariant (PMI) model exhibit a richer phenomenology than linear electrodynamics, allowing first-order phase transitions in both canonical and grand-canonical ensembles, while recovering the Maxwell limit for $p=1$; the corresponding Einstein--PMI black-hole solutions and their thermodynamic, geometric, and holographic properties have been extensively analyzed~\cite{Hassaine:2007py,Hendi:2009zzb,Hendi:2009zza,Hassaine:2008pw,Maeda:2008ha,Hendi:2010zza,Hendi:2010bk,Hendi:2010zz,Hendi:2010kv,Jing:2011vz,Roychowdhury:2012vj}.

Our aim is to show that dynamical instability, quantified by the Lyapunov exponent of unstable circular geodesics, encodes the thermodynamic phase structure induced by nonlinear electrodynamics. We demonstrate that the Lyapunov exponent reproduces the first-order small/large black-hole transition in the canonical ensemble and the second-order critical behavior at its endpoint, becoming multivalued in the coexistence region and exhibiting a finite jump that vanishes at criticality with a universal square-root scaling and critical exponent $1/2$, for both massless and massive probes. Moreover, singular features in its thermal profile coincide with divergences of the heat capacity at fixed charge, establishing a direct link between Lyapunov instability, spinodal behavior, and metastability, and providing a unified dynamical interpretation of black-hole phase transitions; we also clarify that, unlike the Lyapunov exponent, geometric probes such as the critical impact parameter are intrinsically defined only for massless particles.

The paper is organized as follows.
In section \ref{background} we introduce the general formalism of nonlinear electrodynamics coupled to gravity and present the corresponding AdS black-hole solutions for a fixed power-law exponent.
Section \ref{Thermo} is devoted to the thermodynamic analysis and phase structure in the canonical ensemble.
In section \ref{SLE} we compute the Lyapunov exponents associated with unstable circular geodesics.
Section \ref{RLEPT} explores the relationship between the Lyapunov exponent and thermodynamic phase transitions, including both massless and massive probes and the associated critical behavior.
In section \ref{RIP} we investigate the connection between the critical impact parameter and phase transitions, with emphasis on its scaling behavior near criticality.
Section \ref{TSLE} discusses local thermodynamic stability and its relation to the Lyapunov behavior.
Finally, section \ref{conclusion} contains our concluding remarks.


\section{General Formalism for NonLinear Electrodynamics}
\label{background}
 We consider black hole solutions for the power Maxwell theory coupled to gravity described by the action, see Refs. \cite{Hendi:2012um, Zangeneh:2015wia} 
 \begin{eqnarray} \label{action2}
 I = \int \textrm{d}^{4}x\sqrt{-g} \left ( \frac{1}{2 \kappa }  (R-2 \Lambda )+\eta \, |-F_{\mu \nu }F^{\mu \nu}|^p \right)\,,
 \end{eqnarray}
 where $\kappa=8 \pi G=1$, $F_{\mu \nu }=\partial_{\mu} A_{\nu}-\partial_{\nu} A_{\mu}$ and $A_{\mu}$ represents the gauge potential. The exponent $p$ is a rational number and the absolute value ensures that any configuration of electric and magnetic fields can be described by these Lagrangians. One could also consider the Lagrangian without the absolute value and the exponent $p$ restricted to being an integer or a rational number with an odd denominator \cite{Hassaine:2008pw}. The sign of the coupling constant $\eta$ will be chosen such that the energy density of the electromagnetic field is positive. This condition is guaranteed in the following cases: $p > 1/2$ and $\eta > 0$ or $p < 1/2$ and $\eta < 0$ \cite{O.:2016wcf}. Here, we will consider the first condition and a specific value of the exponent $p=3/4$. In Refs.  \cite{Zangeneh:2015wia, Hendi:2012um, O.:2016wcf}  was shown that the field equations have as solutions the topological nonlinearly charged black holes.
The general AdS solution for power Maxwell black hole 
can be written as follow 
\begin{equation}
   ds^2=-f(r)dt^2+\frac{dr^2}{f(r)}+r^2d\Omega^2 \,,
\end{equation}
where $d\Omega^2$ is the metric of the spatial 2-section, which can have a positive ($k=1$), negative ($k=-1$) or zero ($k=0$) curvature, and
\begin{equation} \label{NLPM}
   f(r)= k-\frac{m}{r}- \frac{\Lambda}{3}  r^2+\eta \frac{ 2^{p} (2 p-1)^2}{(3-2p)}\frac{q^{2p}}{r^{\frac{2}{2p-1}}} \,,    
\end{equation}
where $m$ and $q$ are integration constants related to the ADM mass $M$ and the electric charge $Q$ of the black hole by 
\begin{eqnarray}
M&=&4\pi m\,, \\
Q&=& \eta  2^{2 p-1} p \, q^{2 p-1}\,,
\end{eqnarray}
respectively, and the gauge potential is given by
\begin{equation}
   A_t(r) = \frac{1-2p}{3-2p}\, q \,  r^{\frac{3-2p}{1-2p}} \,.
\end{equation}

In Ref. \cite{O.:2016wcf} it has been shown that Eq. (\ref{NLPM}) describes a black hole solution with an inner horizon ($r_{-}$) and an  outer  horizon ($r_+$). The outer horizon $r_+$ of this black hole can be calculated numerically by finding the largest real positive root of $f(r = r_+) = 0$, for different values of parameter $p$ under consideration. 
In the  following, we focus our attention on the specific value $p=3/4$ and $k=1$.
Then, the metric (\ref{NLPM}) reads as follows 
\begin{equation}
f(r)=1 - \frac{M}{4\pi r} + \frac{16 Q^3\, 2^{1/4}}{81 r^4\, \eta^2} + \frac{r^2}{l^{2}} \,. \label{p34}
\end{equation}
The radius of the event horizon $r_h$ is the solution of $f(r_h) = 0$ , then the mass $M$ yields
\begin{equation}
M=4\pi r_h \left(1 + \frac{16\, 2^{\frac{1}{4}}Q^{3}}{81 r_{h}^{4}\eta^{2}} + \frac{r_h^2}{l^{2}}\right) \,.
  \end{equation}

\section{Thermodynamics}
\label{Thermo}

In this section, we review the thermodynamic properties and phase structure of four-dimensional Einstein--Power--Maxwell AdS black holes in the canonical ensemble. Our analysis closely follows the framework developed in Ref.~\cite{Becar:2020mtz}, where the thermodynamics of nonlinear power-law electrodynamics coupled to gravity was systematically studied. For completeness and clarity, we summarize the main thermodynamic quantities and phase-transition features that will be relevant for our subsequent dynamical analysis. Thus, using the surface gravity relation and  Eq. (\ref{p34})  the Hawking temperature of the black hole solutions is given by 
\begin{equation} \label{HTemp}
     T=\frac{1}{4\pi r_h} + \frac{3 r_h}{4\pi l^{2}} - \frac{4\, 2^{1/4} Q^3}{27\pi\, r_h^5\, \eta^2} \,.
\end{equation}
Furthermore, the electric potential $\Phi$, measured at infinity with respect to the horizon, is 
\begin{equation}
\Phi=-(A_t(r_h)-A_t(\infty))=\frac{8 Q^{2}}{27 \eta^{2} r_h^{3}}\,,
\end{equation}
while the Bekenstein-Hawking entropy $S$ is given by
\begin{equation}
\label{entropy}
S=8\pi^{2} r_{h}^2\,.
\end{equation}
In the canonical ensemble (fixed $Q$), the Helmholtz 
free energy can be obtained by evaluating the on-shell Euclidean action. To do so, one must employ the counterterm method to cancel the divergences that arise in the on-shell Euclidean action. Accordingly, in our setup the free energy $F$ is given by
\begin{equation}
 F=M-TS=\pi\left(2r_{h}-\frac{2r_{h}^{3}}{l^{2}}+\frac{160\ 2^{\frac{1}{4}}Q^{3}}{81 r_{h}^{3}\eta^{2}}\right) \,.
\end{equation}\label{escalado}
We now apply dimensional analysis and rescale the following quantities as
\begin{eqnarray} \label{scale}
\notag \tilde{r}_h &=& \frac{r_h}{l} \,, \quad \tilde{F}=\frac{F
}{l} \,, \quad \tilde{T}=T l  \,,\\
\tilde{M} &=&\frac{M}{l} \,, \quad \tilde{r}=\frac{r}{l} \,, \quad \tilde{Q}=\frac{Q}{l^{2/3}} \,,\quad \tilde{\eta}=\eta l \,,
\end{eqnarray}
where the tilde symbol is used here to denote dimensionless quantities.

We now review the phase structure in the canonical ensemble, where the electric charge is held fixed as an extensive thermodynamic parameter. The properties of the black-hole solutions are encoded in the critical behavior of the $(\tilde{T},\tilde{r}_h)$ plane when the rescaled charge $\tilde{Q}$ reaches its critical value $\tilde{Q}_c$. For a given value of $\eta$, the critical point is determined by the simultaneous conditions
\begin{equation}
 \frac{\partial {\tilde{T}}}{\partial{\tilde{r}_{h}}}= \frac{\partial^{2} {\tilde{T}}}{\partial{\tilde{r}_h^{2}}}=0 \,.
\end{equation}
These conditions yield the following critical quantities:
\begin{equation}\label{valorescriticos}
  \tilde{r}_ {hc}=\frac{\sqrt{2}}{3} \,, \quad \tilde{Q}_c= 0.167169 \tilde{\eta}^{\frac{2}{3}} \,, \quad \tilde{T}_c=\frac{3\sqrt{2}}{5 \pi}.
\end{equation}

The temperature as a function of the horizon radius for different values of the charge $\tilde{Q}$, is shown in Fig. 1. This diagram reveals the existence of distinct thermodynamic behaviors depending on whether the charge lies above or below the critical value $\tilde{Q}_c$. For $\tilde{Q} > \tilde{Q}_c$ (illustrated by the black dotted curve), the temperature varies monotonically with $\tilde{r}_h$, indicating the presence of a unique black hole solution for each temperature. In contrast, when $\tilde{Q} < \tilde{Q}_c$ (shown by the blue and red curves), the $\tilde{T}$-$\tilde{r}_h$ curve develops a characteristic swallow-tail structure with two local extrema. This non-monotonic behavior implies that for a given temperature is possible that
three distinct black hole solutions coexist: a small black hole (SBH), an intermediate black hole (IBH), and a large black hole (LBH). Physically, the critical charge  $\tilde{Q}_c$ represents the threshold value that controls the onset and termination of the first-order small/large black-hole phase transition. At $\tilde{Q}=\tilde{Q}_c$, the coexistence region collapses into a second-order critical point, beyond which the system remains in a single thermodynamically stable phase.
\begin{figure}[H]
\centering
\includegraphics[width=0.47\textwidth]{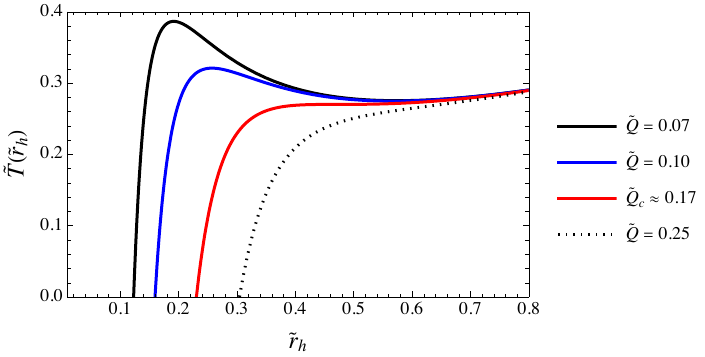}
\caption{Hawking temperature as a function of the horizon radius for $\tilde{\eta}=1$ and for different values of $\tilde{Q}$  below (blue and black) and  above (black dotted) the critical point $\tilde{Q}_{c}$.}
\label{plots0}
\end{figure}

\begin{figure}[H]
\centering
\includegraphics[width=0.4067\textwidth]{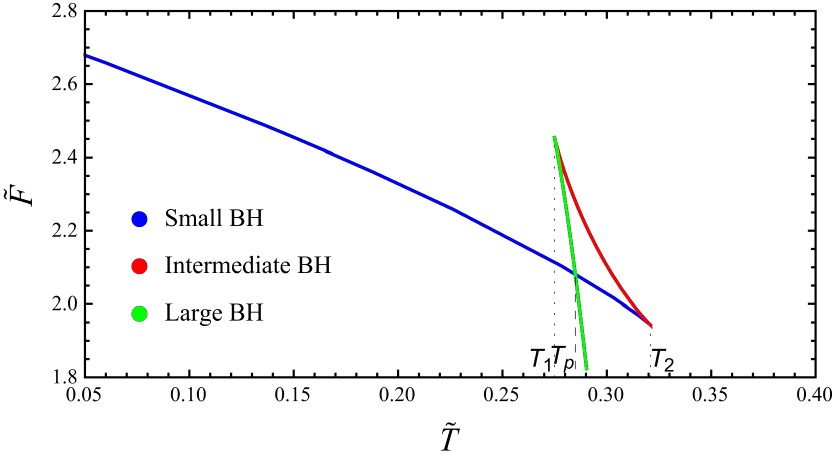}
\includegraphics[width=0.4\textwidth]{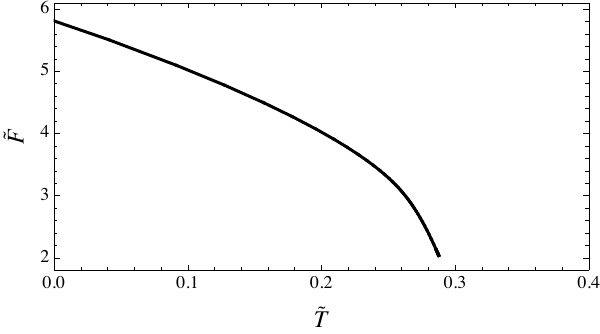}
\caption{The free energy as a function of $\tilde{T}$ for $\tilde{\eta}=1.0$ and different values of $Q$. The top panel corresponds to the case $\tilde{Q}=0.1<\tilde{Q}_c$, while the bottom panel corresponds to $\tilde{Q}=0.25>\tilde{Q}_c$.}
  \label{Potential}
\end{figure}

Now, to study the phase transition, we use the free energy given by Eq. (\ref{escalado}). The rescaled free energy $\tilde{F}$ as a function of $\tilde{T}$ and $\tilde{Q}$ is shown in Fig. \ref{Potential}. When $\tilde{Q}$ is smaller than $\tilde{Q}_{c}$ (top panel) we have three black hole solutions, namely small BH, intermediate BH and large BH. These three black hole solutions can coexist for $\tilde{T}_1 < \tilde{T} < \tilde{T}_{2}$, where $\tilde{T}_1$  and   $\tilde{T}_2$ are the temperatures at points 1 and 2, respectively. The point $p$, where the blue and green curves intersect, corresponds to the first-order phase transition, occurring at $\tilde{T}_p \approx 0.2978$. When $\tilde{Q}$ is greater than $\tilde{Q}_{c}$ (bottom panel), there will be no phase transition as we have only a single black hole solution.

\section{Lyapunov exponents}
\label{SLE}

In this section, we compute the Lyapunov exponents of massless and
massive particles of four-dimensional AdS black holes with a nonlinear electrodynamics source. In particular, we focus on unstable circular geodesics in the equatorial hyperplane with $\theta=\frac{\pi}{2}$. We emphasize that, in general, nonlinear electrodynamics leads to photon propagation along null geodesics of an effective optical metric rather than those of the background geometry \cite{Novello:1999pg}. In the present case, however, the null geodesic structure remains unchanged, in contrast to AdS Born–Infeld black holes \cite{Yang:2023hci}; this result is proven in the Appendix \ref{EMPMI}.
To proceed with the geodesic analysis, we start from the point-particle Lagrangian on the equatorial plane:
\begin{equation}
2L = -f(r) \dot{t}^2 + \frac{1}{f(r)} \dot{r}^2 + r^2 \dot{\phi}^2 \,.
\end{equation}
Here, dots 
denote derivatives with respect to the affine parameter. 
Since $t$ and $\phi$ are cyclic coordinates of the Lagrangian, their associated generalized momenta are conserved, leading to
\begin{equation}
    \dot{\phi}=\frac{L}{r^{2}} \,,
\end{equation}
\begin{equation}\label{tiempo}
    \dot{t}=\frac{E}{f(r)}\,.
\end{equation}
Then, the radial motion can be expressed as
\begin{equation}
\dot{r}^2 = E^2- V_{\text{eff}}(r)\equiv V_{r}(r)
\end{equation}
 where the constant $E$ can be treated as the energy and the energy per unit mass for massless and massive particles, respectively. Here, we introduce the effective potential

 \begin{equation}
V_{\text{eff}}(r) = f(r) \left( \frac{L^2}{r^2} + \epsilon \right) \,,
 \end{equation}
 where $L$ is identified as the angular momentum of the particles, and $\epsilon$ = 0 and 1 correspond to massless and massive particles, respectively. The radius of an unstable circular geodesic ($r_0$) is determined by the condition
 \begin{equation}
     V_{\text{eff}}'(r_0) = 0, \quad V_{\text{eff}}''(r_{0}) < 0 \,,
 \end{equation}
where primes denote derivatives with respect to $r$. The Lyapunov exponents quantify the average exponential rate at which nearby trajectories in phase space diverge or converge over time. For a dynamical system, they provide a measure of sensitivity to initial conditions: positive exponents indicate divergence (and often chaos), while negative exponents indicate convergence toward stable behavior, and zero exponents correspond to neutral stability such as periodic motion.

To compute Lyapunov exponents, one typically linearizes the dynamics along a trajectory using the Jacobian matrix \cite{Cardoso:2008bp,Pradhan:2012rkk,Pradhan:2013bli}, which contains the partial derivatives of the equations of motion with respect to the phase space variables. Taking into account the two-dimensional phase space $(r,\pi_r)$, the Jacobian matrix $K_{ij}$ is defined as
\begin{eqnarray}
    K_{11}= \frac{\partial F_1}{\partial r}\,, \quad  K_{12}= \frac{\partial F_1}{\partial \pi_r}\,, \\
    K_{21}= \frac{\partial F_2}{\partial r}\,, \quad K_{22}= \frac{\partial F_2}{\partial \pi_r}\,,
\end{eqnarray}
where $F_1(r,\pi_r)=\frac{dr}{dt}$, and $F_2(r,\pi_r)=\frac{d\pi_r}{dt}$. When circular motion of particles is considered, $\pi_r=0$, the Jacobian matrix can be reduced to 
 \begin{equation}
     K_{ij}= \begin{pmatrix}
0 & K_{12}\\
K_{21} & 0 
\end{pmatrix} \,,
 \end{equation}
and the eigenvalues of the Jacobian matrix, i.e., the Lyapunov exponents $\lambda$ are given by
\begin{equation}
    \lambda= \pm \sqrt{K_{12}K_{21}} \,.
\end{equation}
If $\lambda^2>0$ the circular motion is unstable, if $\lambda^2=0$ the circular motion is marginal, and if $\lambda^2<0$ the circular motion is stable. So, for circular orbits we have  $\dot{r}=0\Rightarrow V_r=V'_r=0$, and
the Lyapunov exponents are
\begin{equation}
\label{sle}
    \begin{aligned}
        \lambda = \pm \sqrt{\frac{V_r''}{2\dot{t}^2}}~.
    \end{aligned}
\end{equation}

\section{Relationship between Lyapunov exponent and phase transition}
\label{RLEPT}

In this section, we investigate the relationship between Lyapunov exponents of both massless and
massive particles and the phase transitions of four-dimensional AdS black holes with a nonlinear electrodynamic source.

 \subsection{Massless particles}
 For circular null geodesics, the Lyapunov exponent is obtained from the second derivative of the effective radial potential evaluated at the radius of the unstable
photon orbit. Using the general definition of $\lambda$ for a radial perturbation around a circular trajectory, the Lyapunov exponent for massless particles takes the
form
\begin{eqnarray}
\label{nomasivo}
\notag    \lambda
    &=&
    \sqrt{
    \frac{f(r_0)^2  \, V_{r}^{\prime\prime}(r_0)}
         {2 \, E^{2}}
    }, \\
\notag    &=& r_0 \sqrt{f(r_0)} \bigg( \frac{6 r_h + 6 r_h^3 /l^2 + 32 \times 2^{1/4} Q^3 /(27 \eta^2 r_h^3)}{r_0^5}\\
&&  -\frac{3}{r_0^4} -\frac{112 \times 2^{1/4} Q^3}{27 \eta^2 r_0^8} \bigg)^{1/2} \,,
\end{eqnarray}
where \(r_0\) denotes the radius of the null circular geodesic, \(L\) is the angular momentum of the photon, 
and \(V_r(r)\) is the corresponding effective radial potential.  
The condition \(V_{r}(r_0)=V_{r}^{\prime}(r_0)=0\) determines the circular photon orbit, while $V_{r}^{\prime\prime}(r_0)>0$ signals its instability and ensures that \(\lambda\) is real and positive.

\begin{figure}[H]
\centering
\includegraphics[width=0.4\textwidth]{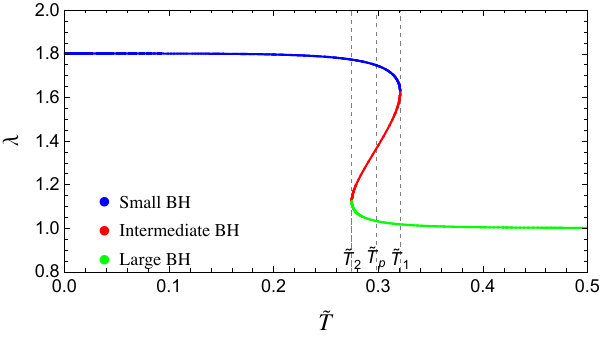}
\includegraphics[width=0.4\textwidth]{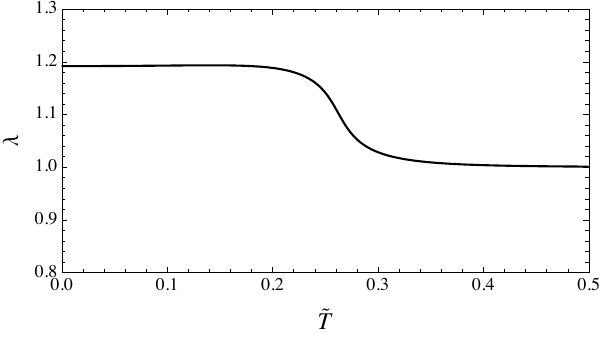}
\caption{ Lyapunov exponent $\lambda$ of the massless particle as a function of temperature $\tilde{T}$ for $\tilde{\eta}=1.0$ and different values of $\tilde{Q}$. Top panel corresponds to $\tilde{Q}=0.1<\tilde{Q}_c$ and the bottom panel to $\tilde{Q}=0.25>\tilde{Q}_c$.}
  \label{Lyapunov}
\end{figure}

We now compute the thermal profile of the Lyapunov exponent associated with massless particles moving in the black hole background. 
The Lyapunov exponent as a function of black hole temperature is shown in Fig. \ref{Lyapunov}, for $\tilde{Q}<\tilde{Q}_{c}$ (top panel), the Lyapunov exponent exhibits multi-valuedness in some temperature range. As $\tilde{T}$ increases from 0 to $\tilde{T}_{1}$, $\lambda$ gradually decreases, as shown by the blue curve. 
At $\tilde{T}_{1}$, the Lyapunov exponent becomes multi-valued.
In particular, in the temperature range $\tilde{T}_{2}\leq \tilde{T}\leq \tilde{T}_{1}$, there are two profiles for which
the Lyapunov exponent decreases with increasing temperature (blue and green curves) and one profile for which it increases with temperature (red curve). The former two profiles appear in the small and large black hole phases,
whereas the latter profile appears in the intermediate phase. The Lyapunov exponents in
the small, intermediate, and large black hole phases are represented by blue, red, and green
lines, respectively, in Fig. \ref{Lyapunov}. 
The dashed vertical lines indicate the spinodal temperatures $\tilde T_{1,2}$, where $(\partial \tilde T/\partial r_h)_Q=0$ and thus the heat capacity $C_Q$ diverges, as well as the coexistence temperature $\tilde T_p$, as will be shown in Sec.~\ref{sec:seven}. 
The multi-valued nature of $\lambda$ corresponds to the swallow tail in the $\tilde{F}-\tilde{T}$ diagram. In addition, we have also shown the small/large black hole transition temperature $\tilde{T}_{p}\approx 0.2978$.\\

The Lyapunov exponent becomes single-valued when $\tilde{Q} > \tilde{Q}_{c}$,  (see Fig. \ref{Lyapunov}, bottom panel).  At low temperatures, $\lambda$ remains approximately constant at a relatively large value, indicating strong geodesic instability associated with the small--black-hole branch. As $\tilde{T}$ increases, $\lambda$ undergoes a rapid but continuous decrease within a narrow temperature interval, smoothly interpolating between two plateau values. At higher temperatures, the Lyapunov exponent stabilizes at a lower value, corresponding to the large--black-hole branch. The absence of multivalued behavior or finite discontinuities indicates that the system lies outside the first-order phase-transition region. The smooth crossover profile of $\lambda(\tilde{T})$ is consistent with a regime close to or above the critical charge $Q_c$, where the small/large black-hole coexistence terminates and the thermodynamic phase structure is characterized by a single stable phase. Finally, it is worth to mentioning that at the critical charge $\tilde{Q}=\tilde{Q}_c$, the three branches merge into a single smooth curve, the discontinuity vanishes, and the system undergoes a second-order phase transition.

\subsection{Massive particles}

\begin{figure}[H]
\centering
\includegraphics[width=0.45\textwidth]{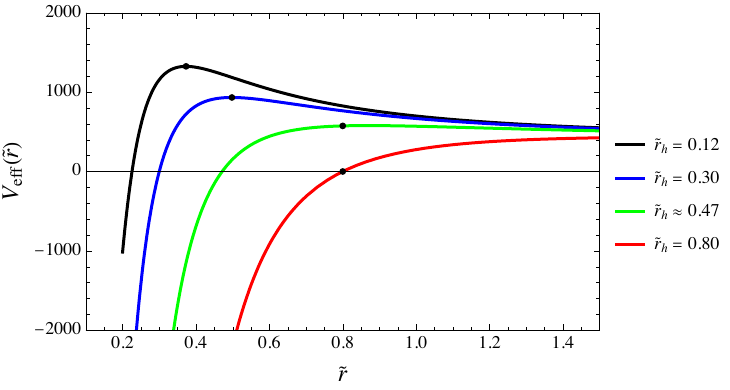}
\caption{ Effective potential $V_{\text{eff}}(\tilde{r})$ of timelike geodesics as a function of $\tilde{r}$ with $\tilde{Q}=0.1$; $\tilde{\eta}=1.0$. The black dots represent the maxima of the effective potentials, corresponding to unstable time-like circular geodesics. For $\tilde{r}_h=0.8$, the effective potential has no maximum.}
\label{MASSIVE}
\end{figure}

For timelike geodesics, we chose $\epsilon = 1$. Conditions $V_{r}(r_{0}) = V_{r}'(r_{0}) = 0$ and $V_{r}''(r_{0}) < 0$ for unstable geodesics provide the following relations for the energy and the angular momentum
\begin{equation}\label{Energy}
    E^{2}=\frac{2 f(r_{0})^{2}}{2 f(r_0)-r_{0} f'(r_0)} \,,
\end{equation}
\begin{equation}\label{angular}
    L^{2}=\frac{r^{3}f'(r_0)}{2 f(r_0)-r_{0}f'(r_0)} \,,
\end{equation}
which requires
\begin{equation}
    2 f(r_0)-rf'(r_0)>0 \,,
\end{equation}
since the energy must be real, here \( r_0 \) denotes the radius of the unstable circular geodesic. The effective potential \( V_{\text{eff}}(r) \) corresponding to unstable timelike geodesics is depicted in Fig.~\ref{MASSIVE} for various values of \( \tilde{r}_{h} \), with \( \tilde{Q} = 0.1 \) and \( L = 20 \). For values of \( \tilde{r}_h > 0.8 \), the effective potential does not exhibit extrema for massive particles. Therefore, this region will not be considered in our subsequent analysis of the Lyapunov exponents.

\begin{figure}[H]
\centering
\includegraphics[width=0.4\textwidth]{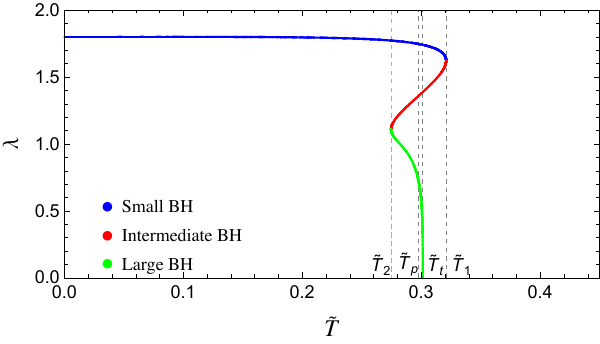}
 \includegraphics[width=0.4\textwidth]{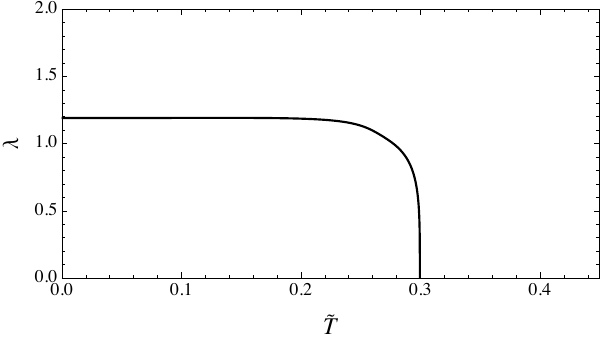}
\caption{Lyapunov exponents $\lambda$ of massive particles as a function of the temperature $\tilde{T}$. Top panel for $\tilde{Q}=0.1<\tilde{Q}_c$ and bottom panel for $\tilde{Q}=0.25>\tilde{Q}_c$.}
\label{massiveLYapu}
\end{figure}

Applying the timelike geodesic conditions (\ref{tiempo}) and (\ref{Energy}) in (\ref{sle}) allows us to obtain the Lyapunov exponents
\begin{eqnarray}\label{masivo}
\notag   \lambda &=&  \frac{1}{2} \sqrt{-(2 f(r_0)-r_0 f^{'}(r_0))V^{''}_r(r_0)} \,.  \\
\notag   &=& \frac{f(r_0)}{9 r_0^4 } \Big[ \frac{1}{E^2} \Big( -\frac{81 r_0^8}{\ell^2}+\frac{81 r_0^5 r_h^3}{\ell^2}+\frac{16 \times 2^{1/4} Q^3 r_0^5}{\eta ^2 r_h^3}- \\  
\notag    && \frac{160 \times 2^{1/4}  Q^3 r_0^2}{\eta ^2}+81 r_0^5 r_h    \Big)  +\frac{L^2}{E^2} \Big( \frac{486 r_0^3 r_h^3}{\ell^2}+ \\    
  && \frac{48 \times 2^{1/4} Q^3 \left(2 r_0^3-7 r_h^3\right)}{\eta ^2 r_h^3}-243 r_0^3 (r_0-2 r_h) \Big)     \Big]^{1/2}
\end{eqnarray}
The Lyapunov exponent as a function of temperature for $L=20$ is shown in Fig. \ref{massiveLYapu}. The top
panel shows $\lambda$ for $\tilde{Q}=0.1$, which is below the critical value $\tilde{Q}_c$. For this value of $\tilde{Q}$, the Lyapunov exponent is multivalued and has three branches. These branches correspond to three different phases or three black hole solutions which can coexist for $\tilde{T}_2<\tilde{T}<\tilde{T}_t$. The phase transition from small black hole to large black hole occurs at the temperature  $\tilde{T}_p\approx 0.2978$. On the other hand, in the bottom panel, for $\tilde{Q}=0.25>\tilde{Q}_c$, there is no phase transition and the Lyapunov exponent is found to be single valued, demonstrating that there is only one black hole solution. Unlike the case of massless particles, for massive particles $\lambda$ vanishes at the temperature point $\tilde{T}=\tilde{T}_t$.

\subsection{Critical Exponent at the Small/Large Transition  from Lyapunov Analysis}
\begin{figure}[H]
\centering
\includegraphics[width=0.4\textwidth]{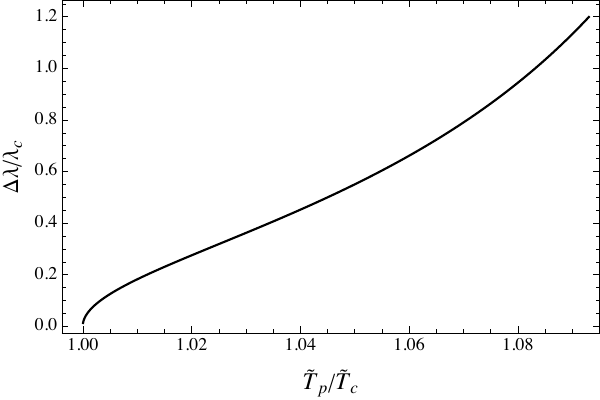}
\caption{ $\Delta \lambda / \lambda_c$ vs $ \tilde{T}_p/ \tilde{T}_c$  curve for null geodesics with $\lambda_c=1.18691$.}\label{Diferencia_nomasivo}
\end{figure}

\begin{figure}[H]
\centering
\includegraphics[width=0.4\textwidth]{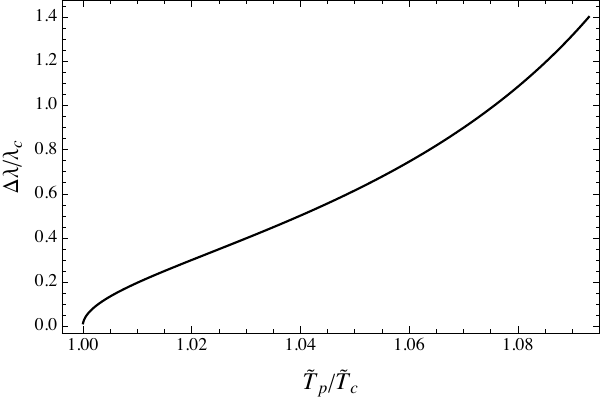}
\caption{ $\Delta \lambda / \lambda_c$ vs $ \tilde{T}_p/ \tilde{T}_c$  curve for timelike geodesics with $\lambda_c=1.17572$.}\label{Diferenciamasivo}
\end{figure}

Now, we study the difference of Lyapunov exponents in the phase transition point 
for massless and massive particles. At the small/large phase transition point $p$, the Lyapunov exponent for small and large black
holes is denoted as $\lambda_{s}$ and $\lambda_{l}$, respectively. With different values of $\tilde{Q}<\tilde{Q}_c$, the phase transition temperature $\tilde{T_{p}}$
changes and for these values we calculate the difference of Lyapunov exponents $\Delta\lambda = \lambda_{s}-\lambda_{l}$.
At the critical point $\tilde{Q}=\tilde{Q}_{c}$, the two extreme points of 
$\tilde{T}$ vs $\tilde{r}_{h}$ curve coincide, and $\tilde{T}_{p} =\tilde{T}_{c}$ and $\lambda_s = \lambda_l = \lambda_c$, which yields $\Delta\lambda = 0$. The critical value of the Lyapunov
exponent $\lambda_c$ can be calculated by inserting the critical values given in Eq. (\ref{valorescriticos}) and it is found as $\lambda_c = 1.18691$ and $\lambda_c = 1.17572$ for massless and massive particles, respectively.
In Figs. \ref{Diferencia_nomasivo} and \ref{Diferenciamasivo} we plot the curve $\Delta\lambda/\lambda_c$ versus $\tilde{T}_p/\tilde{T}_c$. Across the first-order SBH$\to$LBH transition, the Lyapunov exponent displays a discontinuous jump from $\lambda_s$ to $\lambda_l$, yielding a finite $\Delta\lambda$. This supports treating $\Delta\lambda$ as an order parameter for massless particles. On the other hand, strictly speaking, the Lyapunov exponent for massive probes is not a thermodynamic order parameter, since it is probe dependent and, in particular, depends on the chosen conserved energy $E$ or angular momenta $L$. 
Nevertheless, once $E$ or  $L$ is fixed, the discontinuity
across the coexistence line behaves as an \emph{order-parameter--like} dynamical quantity: at the first-order small/large black-hole transition it takes a nonzero value, $\Delta\lambda\neq 0$, providing a sharp signature of the phase change. 
In this sense, $\Delta\lambda$ can be regarded as a (probe-conditioned) dynamical order parameter for the phase transition.

To study the critical behavior of $\Delta{\lambda}$ we calculate the critical exponent, a numerical value that characterizes the behavior of a physical system near its critical point. The order-parameter candidate is the discontinuity
\begin{equation}
\Delta\lambda \equiv \lambda_s-\lambda_l,
\end{equation}
whose critical scaling we now derive. Adopting the standard mean-field ansatz
\begin{equation}
\Delta\lambda \;\sim\; \bigl|\tilde T-\tilde T_c\bigr|^{\sigma},
\label{eq:25}
\end{equation}
the exponent $\sigma$ can be determined from a local expansion around the critical point. So, we introduce small deviations of the horizon radius and temperature according to
\begin{align}
\tilde r_h &= \tilde r_{hc}\,\bigl(1+\Delta\bigr), 
\qquad |\Delta|\ll 1, 
\label{eq:26}
\\[2pt]
\tilde T(\tilde r_h) &= \tilde T_c\,\bigl(1+\varepsilon\bigr), 
\qquad |\varepsilon|\ll 1,
\label{eq:27}
\end{align}
and expand the Lyapunov exponent at fixed charge (or ensemble parameters) about $\tilde r_{hc}$
\begin{equation} \label{eq:28}
\lambda(\tilde r_h) 
= \lambda_c 
+  \frac{\partial \lambda}{\partial \tilde r_h}\bigg|_{c}\, \delta \tilde r_h \,+\, \mathcal{O}\bigl(\delta \tilde r_h^{\,2}\bigr) \,.
\end{equation}
Evaluating \eqref{eq:28} on the two coexisting branches, and using $\lambda_s(\tilde r_{hc})=\lambda_l(\tilde r_{hc})=\lambda_c$, we obtain the dimensionless jump
\begin{equation}
\Delta\tilde\lambda \;\equiv\; \frac{\Delta\lambda}{\lambda_c} 
= \frac{\tilde r_{hc}}{\lambda_c}\left.\frac{\partial \lambda}{\partial \tilde r_h}\right|_{c}\!\bigl(\Delta_s-\Delta_l\bigr)\,.
\label{eq:29}
\end{equation}

\medskip
\noindent
Near criticality, the Hawking temperature admits the quadratic Landau expansion at fixed ensemble,
\begin{equation}
\tilde T 
= \tilde T_c \;+\; \frac{\tilde r_{hc}^{\,2}}{2}\left.\frac{\partial^{2}\tilde T}{\partial \tilde r_h^{\,2}}\right|_{c}\!\Delta^{2} \;+\; \mathcal{O}(\Delta^{3}),
\label{eq:30}
\end{equation}
which implies $\Delta \propto \sqrt{\tilde T/\tilde T_c-1}$ along the coexistence line as the critical point is approached.
Combining \eqref{eq:29}–\eqref{eq:30} yields the mean-field square-root law
\begin{equation}
\Delta\tilde\lambda 
= A\,\sqrt{t-1}\,,
\qquad
t:=\frac{\tilde T}{\tilde T_c}\,,
\label{eq:31}
\end{equation}
where the amplitude is fully determined by local derivatives evaluated at criticality,
\begin{equation}
A
= \sqrt{\frac{\tilde T_c}{\lambda_c}}
\left[
\frac{1}{2}\left.\frac{\partial^{2}\tilde T}{\partial \tilde r_h^{\,2}}\right|_{c}
\right]^{-1/2}
\left.\frac{\partial\,\Delta\lambda}{\partial \tilde r_h}\right|_{c}.
\label{eq:32}
\end{equation}
Equation \eqref{eq:31} immediately identifies the critical exponent
as $\sigma=\tfrac12$, in agreement with the universal mean-field behavior characteristic of Van der Waals–type phase transitions.

\begin{figure}[H]
\centering
\includegraphics[width=0.4\textwidth]{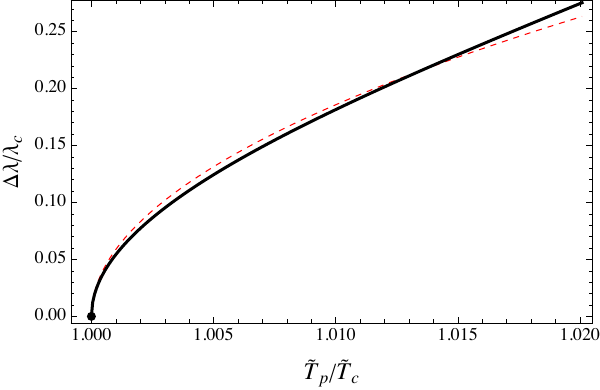}
\caption{ $\Delta \lambda / \lambda_c$ vs $ \tilde{T}_p/ \tilde{T}_c$  curve for null geodesics near critical point (black point).The dashed red line corresponds to $\frac{\Delta\lambda}{\lambda_c}=1.85618\sqrt{\frac{\tilde{T}_p}{\tilde{T}_c}-1}$.}
\label{difference}
\end{figure}

\begin{figure}[H]
\centering
\includegraphics[width=0.4\textwidth]{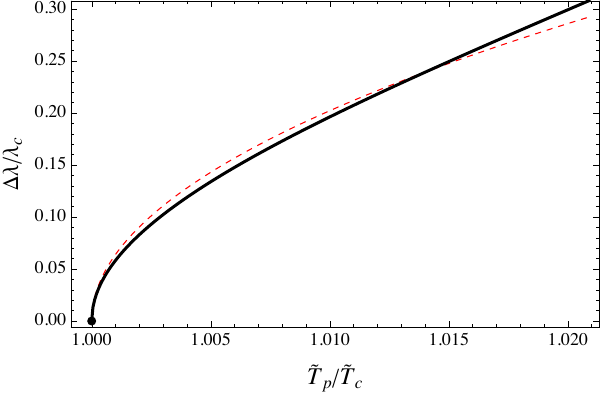}
\caption{ $\Delta \lambda / \lambda_c$ vs $ \tilde{T}_p/ \tilde{T}_c$  curve for time geodesics near critical point (black point).The dashed red line corresponds to $\frac{\Delta\lambda}{\lambda_c}=2.02208\sqrt{\frac{\tilde{T}_p}{\tilde{T}_c}-1}$.}
\label{difference1}
\end{figure}

\medskip
\noindent
For the present model, a numerical evaluation of~\eqref{eq:32} yields the following
square-root fits along the small/large black-hole coexistence line.
For massless particles (null circular orbits),
\begin{equation}
\frac{\Delta\lambda}{\lambda_c}
= 1.85618\,\sqrt{\frac{\tilde T_p}{\tilde T_c}-1}\,,
\label{eq:33}
\end{equation}
whereas for massive particles (timelike circular orbits),
\begin{equation}
\frac{\Delta\lambda}{\lambda_c}
= 2.02208\,\sqrt{\frac{\tilde T_p}{\tilde T_c}-1}\,.
\label{eq:34}
\end{equation}
Both relations capture the approach to criticality with the expected mean-field
exponent \(1/2\), providing model-dependent amplitudes for the Lyapunov jump.
The plot of the rescaled order parameter $\Delta\lambda/\lambda_c$ vs $\tilde{T}_p/\tilde{T}_c$ for the massless and massive particles near the critical point is shown in Figs. \ref{difference} and \ref{difference1}, respectively. The dashed line represents the curve fitting and the solid line shows the data points. 

We can observe that the results fit well  with the analytical curve, Eq. (\ref{eq:31}), near the critical point. Although the value of $A$ can differ, the critical exponent $\sigma$ is always one-half for the black hole, independently of the nature of the particles. Therefore, $\Delta\lambda$ can be considered another order parameter to investigate the first order phase transition between small and large phases of the four-dimensional AdS black holes with a nonlinear electrodynamics source.

\section{Relationship between the critical impact parameter and phase transitions}
\label{RIP}

\subsection{Relationship between the Lyapunov exponent and the critical impact parameter}

Unstable circular null geodesics encode both dynamical and optical information of the black hole geometry. 
Two quantities associated with these orbits are the Lyapunov exponent $\lambda$, which measures the rate of exponential divergence of radial perturbations, and the critical impact parameter $b_c$, which determines the threshold between capture and scattering. Both quantities depend on the same geometric structure at the radius $r_0$ of the unstable photon orbit.

For null geodesics the critical impact parameter follows from the circular-orbit condition and is given by
\begin{equation}
    b_c = \frac{r_0}{\sqrt{f(r_0)}} .
\end{equation}
So, by considering the Lyapunov exponent Eq.  (\ref{nomasivo}), 
one obtains a direct relation between the Lyapunov exponent and the critical impact parameter, which is given by
\begin{equation}
\lambda\, b_c
= r_0^2 
\sqrt{
- \frac{1}
     {2} \frac{d^2}{dr^2} \left( \frac{f(r)}{r^2} \right) \bigg|_{r=r_0}
     }\,.
\label{lambda-bc-relation-null}
\end{equation}

\begin{figure}[H]
\centering
\includegraphics[width=0.45\textwidth]{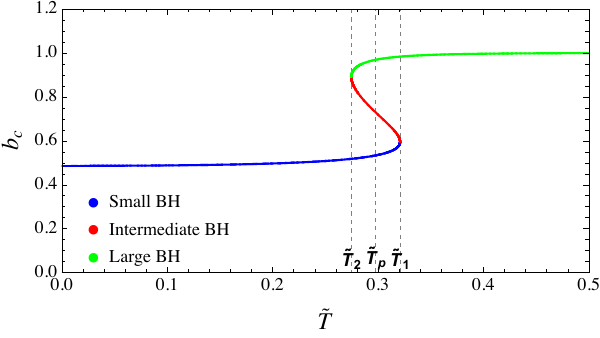}
\caption{The critical parameter  $b_c$ of massless particles as a function of the temperature $\tilde{T}$, for  $\tilde{\eta}=1.0$, and $\tilde{Q}=0.1<\tilde{Q}_c$.}
\label{bc}
\end{figure}

This expression makes explicit that both $\lambda$ and $b_c$ are controlled by the same geometric data 
$\{r_0, f(r_0), f^{\prime}(r_0), f^{\prime\prime}(r_0)\}$; therefore, inherit the same multivalued 
behavior across the small/large black-hole phase transition, which is shown in Fig. \ref{bc}.

On the other hand, for massive probes, a universal critical impact parameter is not available, since the capture-return threshold depends explicitly on the conserved energy $E$. 
For a fixed value of $E$, the critical trajectory corresponds to an unstable circular timelike orbit at $r=r_0$.
It is then convenient to introduce the energy-dependent ratio
\begin{equation}
b_m(E)\equiv \frac{L_c(E)}{E},
\end{equation}
which, using Eqs.~\eqref{Energy} and \eqref{angular} evaluated at $r=r_0$, can be written in parametric form as
\begin{equation}
b_m(E)\equiv \sqrt{\frac{r_{0}^{3} f'(r_0)}{2 f(r_0)^{2}}}\, .
\label{eq:bm_parametric}
\end{equation}
Here $r_0$ is fixed by the chosen energy through the condition $E^{2}=E_c^{2}(r_0)$, so that $b_m$ is not universal but rather depends on the selected massive probe energy.
Since massive critical trajectories are energy dependent, fixing the conserved energy $E$ uniquely determines the radius $r_0$ of the relevant unstable timelike circular orbit. 
Dear It is therefore convenient to express the Lyapunov exponent in terms of the energy-dependent ratio
$b_m(E)\equiv L_c(E)/E$.
Thus, using the critical-orbit relations at $r=r_0$ to eliminate $E$ and $L$ 
one obtains the compact form
\begin{equation}
\lambda
=
\sqrt{-\frac{f(r_0)(r_0^2-f(r_0)b_m^2)V_r''}{2 r_0}}\,.
\label{eq:lambda_bm}
\end{equation}
However, the energy dependence makes it a less consistent indicator of the thermodynamic phase transition, and for this reason the subsequent analysis focuses primarily on the massless case.

\subsection{Scaling behavior of the minimum impact parameter near criticality}
Along the SBH/LBH coexistence line we adopt as order parameter the dimensionless jump of the critical impact parameter,
\begin{equation}
\Delta b(\tilde T_p)\;\equiv\;
\frac{b_c^{(L)}(\tilde T_p)-b_c^{(S)}(\tilde T_p)}{\,b_c|_{(\tilde T_c,\ldots)}}\,,
\label{eq:OrderParam_b_def}
\end{equation}
where $b_c^2=\sqrt{\frac{3}{2}}\frac{r_{ph}^{2}}{f(r_{ph})}$ and $b_c|_{(\tilde T_c,\ldots)}$ denotes the critical value evaluated at the second–order endpoint. Near criticality we find the mean–field power law
\begin{equation}
\Delta b(\tilde T_p)\;=\;
B\,\Bigl(\frac{\tilde T_p}{\tilde T_c}-1\Bigr)^{\beta}\!,
\qquad \beta=\tfrac{1}{2}\,,
\label{eq:OrderParam_b_scaling}
\end{equation}
which quantitatively captures the vanishing of the discontinuity as $\tilde T_p\to\tilde T_c$.
This result parallels the square–root behavior observed for other photon–sphere diagnostics in AdS
black holes \cite{NaveenaKumara:2019nnt,Sood:2024rfr}.

We are in the area 
Two alternative normalizations are commonly used and are straightforwardly related to
\eqref{eq:OrderParam_b_def}:
(i) the \emph{unnormalized} jump, $\Delta b=b_c^{(L)}-b_c^{(S)}$, fitted directly to a square–root
law in $\tilde T_p/\tilde T_c-1$; and
(ii) the \emph{dynamic} normalization by the instantaneous value, $\Delta b/b_c(\tilde T_p)$.
Both choices differ from \eqref{eq:OrderParam_b_def} by a (near–critical) constant or a slowly varying
factor, hence they merely rescale the fitted amplitude while leaving the critical exponent $\beta$
unchanged. In particular, previous analysis employing reduced photon–sphere quantities (e.g. reduced
impact parameter or reduced photon–sphere radius) reported the same mean–field exponent $1/2$ when
regressed against the reduced temperature along coexistence.

\begin{figure}[H]
\centering
\includegraphics[width=0.4\textwidth]{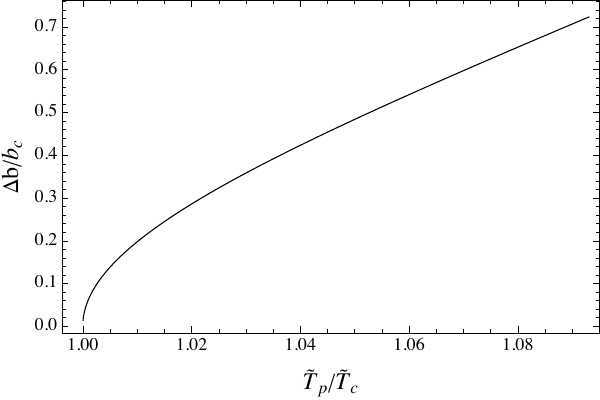}
\caption{ Dimensionless order parameter built from the critical impact parameter, 
$\Delta b \equiv \Delta b/b_c|_{(\tilde T_c,\ldots)}$, along the SBH/LBH coexistence line. 
The curve displays a smooth, monotonic rise from zero at $\tilde T_p/\tilde T_c \to 1$ and follows 
the mean–field square–root law anticipated by the near–critical analysis in Sec.~5.3 and made 
explicit for $\Delta b$ in Sec.~6.2 [Eq. (\ref{eq:OrderParam_b_scaling})]. The vanishing at the endpoint signals the second–order critical point, whereas the overall scale (amplitude) is model–dependent. }
\label{parametro1}
\end{figure}

\begin{figure}[H]
\centering
\includegraphics[width=0.4\textwidth]{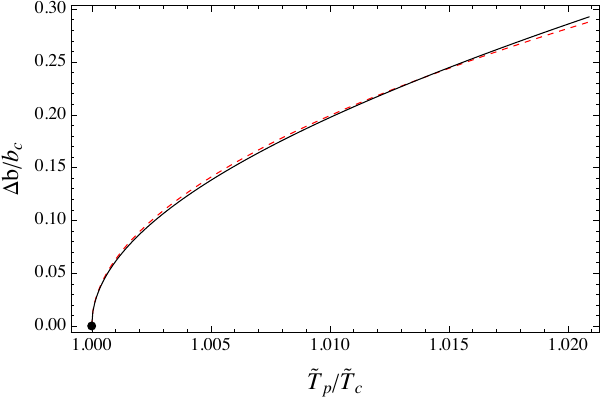}
\caption{Near–critical scaling of the dimensionless order parameter 
$\Delta b \equiv \Delta b/b_c|_{(\tilde T_c,\ldots)}$ versus $\tilde T_p/\tilde T_c$. 
Black points: numerical data; red dashed line: best square–root fit 
$\Delta b=1.99054\,\sqrt{\tilde T_p/\tilde T_c - 1}$ in the near–critical window, in agreement 
with the mean–field prediction $\beta=\tfrac12$ (Sec.~5.3) and the explicit formulation in Sec.~6.2. 
}
\label{difference2} .
\end{figure}

Figs. \ref{parametro1} and \ref{difference2} corroborate the mean–field scenario established in Sec.~5.3: 
the discontinuity of the impact–parameter diagnostic along coexistence, expressed in the 
dimensionless form $\Delta b=\Delta b/b_c|_{(\tilde T_c,\ldots)}$, vanishes continuously at the 
critical point and obeys a square–root law in the reduced temperature, 
Eq.(\ref{eq:OrderParam_b_scaling}). 
The extracted amplitude ($B=1.99054$) is model–dependent, whereas the critical 
exponent $\beta=\tfrac12$ is universal and matches the Lyapunov–exponent analysis of Sec.~5.3. 
Consistent with our general discussion, normalizations based on $\Delta b$ or on 
$\Delta b/b_c(\tilde T_p)$ only change the overall scale of the fit and leave the exponent unchanged.

\section{Local thermodynamic stability and Lyapunov behavior}\label{sec:seven}
\label{TSLE}
In the canonical ensemble, local thermodynamic stability is governed by the heat capacity at fixed charge
\begin{equation}
\label{heat}
C_Q \equiv T\left(\frac{\partial S}{\partial T}\right)_Q,
\end{equation}
where 
the temperature is given by Eq. (\ref{HTemp}) and the entropy by Eq. (\ref{entropy}).
So,
\begin{equation}
\frac{\partial S}{\partial r_h} = 16\pi^2 r_h  \,, \qquad
\frac{\partial T}{\partial r_h}
= \frac{3}{4\pi} - \frac{1}{4\pi r_h^{2}}
+ \frac{20 \times 2^{1/4}Q^{3}}{27\pi\,\eta^{2}\,r_h^{6}} \,.
\end{equation}
Therefore, the heat capacity at fixed charge (\ref{heat}) 
takes the form
\begin{equation}
C_Q = 
T\,\frac{\partial S/\partial r_h}{\partial T/\partial r_h}
= \frac{16\pi^2 r_h\,T(r_h)}{\partial T / \partial r_h} \,.
\label{eq:CQ}
\end{equation}
As a consequence, $C_Q$ diverges whenever
\begin{equation}
\left(\frac{\partial T}{\partial r_h}\right)_Q = 0   \,,
\end{equation}
that is, at the spinodal points where the $T$--$r_h$ curve develops a local extrema. 
These divergences mark the boundaries between locally stable ($C_Q>0$) and unstable ($C_Q<0$) black-hole branches.

Provided that $(\partial\lambda/\partial r_h)_Q$ remains finite and nonvanishing, the spinodal condition $\partial T/\partial r_h=0$  implies a divergence of $(\partial\lambda/\partial T)_Q$. In the parametric representation $(T(r_h),\lambda(r_h))$, this corresponds to the appearance of vertical tangents in the $\lambda-T$ plane. 
Consequently, the spinodal lines at which the heat capacity diverges and changes sign are precisely mirrored in the Lyapunov diagnostic by singular slopes in the thermal profile of $\lambda(T)$.
This correspondence further supports the interpretation of the Lyapunov exponent as a dynamical probe of local thermodynamic stability. Moreover, near the first-order phase transition, where the system bifurcates into small and large black hole branches, the discontionuity of the Lyapunov exponent, $\Delta\lambda=\lambda_s-\lambda_l$, provides a complementary and sharper characterization of the transition, while the divergence of $\left(\frac{\partial \lambda}{\partial T}\right)_Q$ captures the spinodal instability, the nonzero jump $\Delta\lambda$ across the coexistence line encodes the dynamical distinction between the two phases, thereby stablishing a direct link between local thermodynamic instability and the global first-order phase structure.

\begin{figure}[H]
\centering
\includegraphics[width=0.45\textwidth]{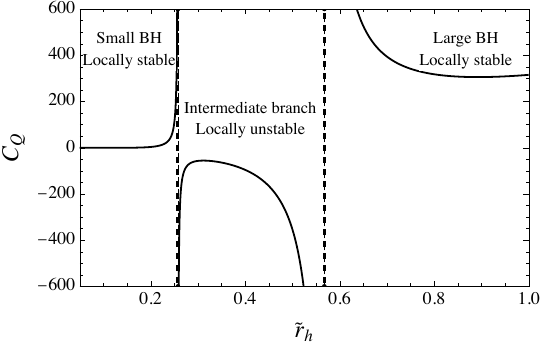}
\caption{Heat capacity at fixed charge $C_Q$ versus the horizon radius $\tilde{r}_h$ 
 with $\tilde{Q}=0.1<\tilde{Q}_c$ and $\tilde{\eta}=1$.}
\label{heatcapacity}
\end{figure}

To make this structure explicit, in Fig.~\ref{heatcapacity} we plot the heat capacity $C_Q(\tilde r_h)$
for $\tilde{\eta}=1$ and $\tilde Q=0.1<\tilde Q_c$. As anticipated from Eq.~(\ref{eq:CQ}),
$C_Q$ diverges at the two spinodal radii $\tilde r_{h1,2}$ determined by
$(\partial \tilde T/\partial \tilde r_h)_Q=0$, i.e. precisely at the local extrema of the
$\tilde T(\tilde r_h)$ curve shown in Fig.~\ref{plots0}. The sign of $C_Q$ then separates the locally stable
small- and large-black-hole branches ($C_Q>0$) from the intermediate unstable branch ($C_Q<0$).
Moreover, since The Lyapunov exponent is a dynamical quantity associated with unstable circular geodesics. 
At fixed $Q$ it can be regarded as a function of the horizon radius through the radius of the circular orbit $\lambda = \lambda(r_0(r_h;Q))$. 
by applying the chain rule, its thermal derivative takes the form
\begin{equation}
\left(\frac{\partial \lambda}{\partial T}\right)_Q
= \frac{\displaystyle \left(\frac{\partial \lambda}{\partial r_h}\right)_Q}
{\displaystyle \left(\frac{\partial T}{\partial r_h}\right)_Q}    \,.
\end{equation}
Therefore, the same spinodal temperatures $\tilde T_{1,2}=\tilde T(\tilde r_{h1,2})$ correspond to vertical tangents
in the $\lambda$--$\tilde T$ profiles, consistently with the dashed lines displayed in Fig.~\ref{Lyapunov}
(massless) and Fig.~\ref{massiveLYapu} (massive).

\section{Concluding Comments}
\label{conclusion}

In this work, we have analyzed the thermodynamic phase structure of four-dimensional Anti--de Sitter black holes in Einstein gravity coupled to a nonlinear power-law electromagnetic field with exponent $p = 3/4$, and investigated how dynamical instability---quantified through the Lyapunov exponent associated with unstable circular geodesics---encodes both first-order and second-order phase transitions in the canonical ensemble.

At fixed electric charge $\tilde{Q}$, we identified a critical value
$\tilde{Q}_c \approx 0.167169\,\tilde{\eta}^{2/3}$,
which separates two qualitatively distinct thermodynamic regimes. For $\tilde{Q} < \tilde{Q}_c$, the system undergoes a first-order small/large black hole (SBH/LBH) phase transition, characterized by a non-monotonic temperature--horizon radius relation with two extrema and the coexistence of three branches: small, intermediate, and large black holes. In this regime, the free energy develops a swallow-tail structure, and the phase transition occurs at a well-defined temperature $\tilde{T}_p$, where the system discontinuously jumps between the SBH and LBH phases. For $\tilde{Q} \geq \tilde{Q}_c$, the swallow tail disappears and the system admits a single thermodynamically stable branch with no phase transition.

This first-order phase transition is faithfully captured by the Lyapunov exponent $\lambda$. In the coexistence region $\tilde{Q} < \tilde{Q}_c$, the thermal profile $\lambda(T)$ becomes multivalued, reflecting the presence of multiple black hole branches. At the transition temperature $T_p$, the Lyapunov exponent exhibits a finite discontinuity,
$\Delta \lambda = \lambda_s - \lambda_l$,
which plays the role of a dynamical order parameter for the SBH/LBH transition. This behavior is observed for both null and timelike unstable circular orbits, indicating that the dynamical signature of the first-order transition is robust and independent of the particle mass.

As the electric charge approaches the critical value $\tilde{Q}_c$, the first-order transition line terminates at a second-order critical point. At this endpoint, the Lyapunov jump vanishes continuously, $\Delta \lambda \to 0$, and the system undergoes a continuous phase transition characterized by divergent response functions. Near criticality, we find that the Lyapunov exponent obeys the universal scaling law
$\Delta \lambda \propto \left( \frac{\tilde{T}_p}{\tilde{T}_c} - 1 \right)^{1/2}$,
corresponding to a critical exponent $\sigma = 1/2$, characteristic of mean-field theory. This result confirms that the SBH/LBH first-order transition terminates at a second-order critical point, in full analogy with the Van der Waals fluid. The same critical behavior is observed for both massless and massive probes and is insensitive to variations of the nonlinear coupling parameter $\eta$.

In addition, our analysis establishes a direct link between dynamical instability and local thermodynamic stability. Singular features in the thermal profile of the Lyapunov exponent coincide with the divergence of the heat capacity at fixed charge, connecting Lyapunov instability with spinodal behavior and metastability. In this sense, the Lyapunov exponent provides a unified dynamical framework that captures both the global features of first-order phase coexistence and the local critical behavior associated with second-order phase transitions.

While geometric probes such as the photon sphere radius and the critical impact parameter have been extensively used to characterize black-hole phase transitions, it is important to emphasize a fundamental distinction between these observables. The critical impact parameter is intrinsically defined only for massless probes, for which the notion of capture versus scattering is well posed and a universal threshold emerges from the unstable circular photon orbit. For massive particles, such a universal impact parameter cannot be defined, as the dynamics depends explicitly on the initial energy and angular momentum and no sharp capture boundary exists. In contrast, the Lyapunov exponent provides a genuinely dynamical probe applicable to both null and timelike unstable circular orbits, highlighting its broader scope and reinforcing its role as the central observable in our analysis.

From a broader perspective, the Lyapunov exponent plays multiple roles in black-hole physics: it sets the instability timescale of circular orbits, governs the damping of quasinormal modes in the eikonal limit \cite{Cardoso:2008bp, Kouniatalis:2025pxs}, controls the angular size of black-hole shadows \cite{Gonzalez:2025yjm}, and satisfies the universal chaos bound
$\lambda \leq \frac{2\pi T}{\hbar}$ \cite{Maldacena:2015waa},
thereby linking gravitational dynamics to quantum thermodynamics. Our results add a new dimension to this picture by demonstrating that Lyapunov instability provides a unified dynamical description of both first-order and second-order black-hole phase transitions.

Finally, we note that although photon propagation in generic nonlinear electrodynamics is governed by an effective optical metric, for the Power--Maxwell Invariant model with $p = 3/4$ and a purely radial electric field the effective metric differs from the background geometry only by trivial rescalings. Consequently, the unstable circular photon orbits and the associated Lyapunov exponent remain unchanged, allowing for a clean dynamical interpretation. Extending this analysis to nonlinear electrodynamics models with genuinely nontrivial optical metrics, as well as to rotating and higher-curvature black holes, constitutes an interesting direction for future work that could further elucidate the universality of Lyapunov instability as a probe of black-hole phase transitions..

\appendix{}

\section{Effective metric and Null Effective Potential in PMI with $p=\tfrac{3}{4}$}
\label{EMPMI}

We consider
a static and spherically symmetric background, given by 
\begin{equation}
ds^2 = -f(r)\,dt^2 + \frac{dr^2}{f(r)} + r^2 d\Omega^2 \,,
\label{eq:background}
\end{equation}
and a purely radial electric field, \(F_{tr}=E(r)\). The nonlinear electrodynamics (NLED) model is given by the Power--Maxwell Invariant (PMI)
\begin{equation}\label{invariant}
L(F) = \eta\,(-F)^p \,, \qquad F \equiv F_{\mu\nu}F^{\mu\nu} \,,
\end{equation}
with signature \((- + + +)\). The electromagnetic potential (Eq.\ (6)) and the electric field are given by 
\begin{equation}
A_t(r) = -\frac{8}{27}\, Q^2\, r^{-3} \,, \qquad
E(r) = \frac{8}{9} Q^{2}\, r^{-4}  \,.
\label{eq:At_E_p34}
\end{equation}
Then, the field invariant Eq.\ref{invariant} can be written as
\begin{equation}
(-F) = \frac{128}{81}\, Q^4\, r^{-8} \,.
\label{eq:F_invariant}
\end{equation}
Due to the nonlinear electrodynamics effects, photons
propagate along null geodesics in the effective metric rather than the background metric. The effective metric $G_{\mu\nu}$ takes the form \cite{Novello:1999pg} 
\begin{equation}
G_{eff}^{\mu\nu} = L_F g^{\mu\nu} - 4 L_{FF}\, F^{\mu}_{\alpha} F^{\nu\alpha}  \,,
\end{equation}
where $L_F=\frac{dL}{dF} $ and $L_{FF}=\frac{d^{2}L}{dF^{2}}$.
Now, defining the standard eikonal combinations
\begin{equation}
\Psi(r) \equiv L_F
= -\frac{9}{8}\, 2^{-3/4}\, Q^{-1/4}\, r^{2} \,,
\label{eq:Psi_def}
\end{equation}
\begin{eqnarray}
\notag  \Phi(r) && \equiv L_F + 4 L_{FF} E^2
= L_F - 2 F L_{FF}\\
&& = -\frac{27\eta}{16}\, 2^{-3/4}\, Q^{-1/4}\, r^{2}  \,,
\label{eq:Phi_def}
\end{eqnarray}
the effective covariant metric for NLED in the eikonal limit can be expressed in the following form
\begin{equation}
ds^2_{\text{eff}}
= -\frac{f(r)}{\Phi(r)}\,dt^2
+ \frac{d r^2}{f(r)\,\Phi(r)}
+ \frac{r^2}{\Psi(r)}\,d\Omega^2  \,,
\label{eq:geff_cov}
\end{equation}
with \(\Phi,\Psi\) from \eqref{eq:Psi_def}--\eqref{eq:Phi_def}. Thus, the metric components are
\begin{eqnarray}
g^{\text{eff}}_{tt}(r) &=& \frac{16\, 2^{3/4}\, Q^{1/4}}{27\,\eta}\, \frac{f(r)}{r^{2}} \,, \\
g^{\text{eff}}_{rr}(r) &=& -\,\frac{16\, 2^{3/4}\, Q^{1/4}}{27\,\eta}\, \frac{1}{f(r)\, r^{2}} \,, \\
g^{\text{eff}}_{\theta\theta}(r) &=& -\,\frac{8\, 2^{3/4}\, Q^{1/4}}{9\,\eta} \,, \\
g^{\text{eff}}_{\phi\phi}(r) &=& g^{\text{eff}}_{\theta\theta}(r)\, \sin^2\theta    \,.
\end{eqnarray}
Therefore, Eq.(\ref{eq:geff_cov}) can be rewritten in the form of a general conformal factor as 
\begin{equation}
ds^2_{eff}
= \Omega^2(r)\, \bigg[-fdt^2 + \frac{dr^2}{f} + h(r)\,d\Omega^2 \bigg],
\label{eq:global_prefactor}
\end{equation}
with $\Omega^2(r) = \frac{1}{\Phi(r)}$ and $h(r)=\frac{\Phi}{\Psi}\, r^2 =\frac{3}{2}\, r^2 $.
Now, considering the equatorial plane $\theta=\pi/2$, the geodesic Lagrangian of the metric \eqref{eq:global_prefactor} takes the following form:
\begin{equation}
\mathcal{L}=\tfrac12\,g^{\mathrm{eff}}_{\mu\nu}\,\dot x^\mu \dot x^\nu
=\tfrac12\,\Omega^2(r)\left[-\,f\,\dot t^2+\frac{\dot r^2}{f}+h\,\dot\phi^2\right] \,,
\label{eq:Lnull}
\end{equation}
and the associated (Killing) constants of motion are
\begin{equation}
E:= -\,\frac{\partial\mathcal{L}}{\partial \dot t}
=\Omega^2 f\,\dot t\,,
\qquad
L:= \frac{\partial\mathcal{L}}{\partial \dot \phi}
=\Omega^2 h\,\dot\phi\,.
\label{eq:EL}
\end{equation}
The null condition \((\mathcal{L}=0)\) from \eqref{eq:Lnull} gives
\begin{equation}
-\,\Omega^2 f\,\dot t^2+\Omega^2\frac{\dot r^2}{f}+\Omega^2 h\,\dot\phi^2=0\,,
\end{equation}
and substituting \(\dot t=E/(\Omega^2 f)\) and \(\dot\phi=L/(\Omega^2 h)\) from \eqref{eq:EL}, we obtain:
\begin{equation}
-\,\frac{E^2}{\Omega^2 f}+\Omega^2\frac{\dot r^2}{f}+ \frac{L^2}{\Omega^2 h}=0
\ \ \Longrightarrow\ \
\dot r^2=\frac{E^2}{\Omega^4}-\frac{f}{\Omega^4}\,\frac{L^2}{h}   \,.
\label{eq:radial_raw}
\end{equation}
Now, defining a rescaled affine parameter \(\sigma\) by \(\mathrm{d}\sigma=\mathrm{d}\lambda/\Omega^2\) (equivalently, \(\dot r=(\mathrm{d}r/\mathrm{d}\sigma)/\Omega^2\)), \eqref{eq:radial_raw} becomes
\begin{equation}
\left(\frac{\mathrm{d}r}{\mathrm{d}\sigma}\right)^2
= E^2 - \frac{f(r)}{h(r)}\,L^2  \,.
\label{eq:radial_sigma}
\end{equation}
Therefore, \emph{all} \(\Omega^2(r)\)-dependence drops out of the null radial equation.
Casting \eqref{eq:radial_sigma} in the standard form
\begin{equation}
\left(\frac{\mathrm{d}r}{\mathrm{d}\sigma}\right)^2+V_{\mathrm{null}}(r)=E^2,
\end{equation}
the effective potential for null particles is identified as
\begin{equation}
V_{\mathrm{null}}(r)=\frac{f(r)}{h(r)}\,L^2
=\frac{f(r)}{h(r)}\left(\frac{L}{E}\right)^2 E^2   \,.
\end{equation}
Hence, \emph{in units of \(E^2\)} (or equivalently, in terms of the impact parameter \(b:=L/E\)),
\begin{equation}
\ V_{\mathrm{null}}(r)\big/E^2 \;=\; \frac{f(r)}{h(r)}\,b^2\  \qquad (\text{independent of } \Omega^2).
\label{eq:Vnull_main}
\end{equation}

Therefore, for PMI with $p=\frac{3}{4}$ and a purely radial electric field, the conformal factor in the effective optical metric cancels out of the null radial equation. Consequently, the null effective potential and the circular photon-orbit structure are determined solely by the background function $f(r)$, unlike in AdS Born–Infeld black holes.

\acknowledgments

Y. V. acknowledges support by the Direcci\'on de Investigaci\'on y Desarrollo de la Universidad de La Serena, Grant No. PR25538511.

\end{document}